\documentclass[journal]{IEEEtran}

\usepackage[cmex10]{amsmath}
\usepackage{amssymb}
\usepackage[mathscr]{eucal}
\usepackage{cite}
\usepackage{amsbsy}
\usepackage{esint}
\usepackage{tikz}
\usepackage{bbm}
\usepackage{graphicx}
\usepackage{caption}
\usepackage{subfigure}
\usepackage{makecell}
\usepackage{multirow}
\usepackage{url}

\newtheorem{remark}{Remark}[section]

                    {$\blacksquare$\vspace*{7pt}} % square <--> blacksquare

\def\s{\sigma}

\def\bi{{\mathbf i}}

\def\x{\boldsymbol{x}}

\def\y{{\boldsymbol y}}

\def\I{{\mathbf I}}

\def\bX{\mathbf{X}}
\def\bY {\mathbf{Y}}
\def\bP{\mathbf{P}}

\def\bi{\begin{itemize}} \def\ei{\end{itemize}}
\def\be{\begin{eqnarray*}}
\def\ee{\end{eqnarray*}}

\def\0{{\mathbf 0}}

\newcommand{\beq}{\begin{equation}}
\newcommand{\eeq}{\end{equation}}

\def\eref#1{(\ref{#1})}

\def\XXint#1#2#3{{\setbox0=\hbox{$#1{#2#3}{\int}$ }
\vcenter{\hbox{$#2#3$ }}\kern-.55\wd0}}

\def\CC{{\mathbf{C}}}

\def\b{{\boldsymbol{b}}}

\def\s{{\boldsymbol{s}}}

\ifCLASSINFOpdf
\DeclareMathOperator*{\argmin}{arg\,min}
\else

\fi

\usepackage{algorithmic}

\usepackage{array}

\usepackage{fixltx2e}

\usepackage{stfloats}

\usepackage{url}
\RequirePackage{lineno}

\begin{document}

\title{Automatic evaluation of fetal head biometry from ultrasound images using machine learning}

\author{Hwa Pyung Kim, Sung Min Lee, Ja-Young Kwon$^*$, Yejin Park, Kang Cheol Kim, and Jin Keun Seo
\thanks{Manuscript received XXX; revised  XXX.
	H. P. Kim, S. M. Lee, K. C. Kim, and J. K. Seo are with the Department of Computational Science and Engineering, Yonsei University, Seoul 03722, South Korea. (e-mail: hpkim0512@yonsei.ac.kr; sungminlee@yonsei.ac.kr; kangcheol@yonsei.ac.kr; seoj@yonsei.ac.kr)
	
	J.-Y. Kwon and Y. Park are with the Department of Obstetrics and Gynecology, Institute of Women's Life Medical Science, Yonsei University College of Medicine, Seoul 03722, South Korea. (e-mail: jaykwon@yuhs.ac.kr; dryjpark02@yuhs.ac.kr)
	
	Asterisk indicates the corresponding author (email: jaykwon@yuhs.ac.kr)}
}

\markboth{}%
{Kim \MakeLowercase{\textit{et al.}}: }

\IEEEtitleabstractindextext{%
\begin{abstract}
Ultrasound-based fetal biometric measurements, such as  head circumference (HC) and biparietal diameter (BPD), are commonly used to evaluate the gestational age and diagnose fetal central nervous system (CNS) pathology. Since manual measurements are operator-dependent and time-consuming, there have been numerous researches on automated methods. However, existing automated methods still are not satisfactory in terms of accuracy and reliability, owing to difficulties in dealing with various artifacts in ultrasound images. This paper focuses on fetal head biometry and develops a deep-learning-based method for estimating HC and BPD with a high degree of accuracy and reliability.
The proposed method effectively identifies the head boundary by differentiating tissue image patterns with respect to the ultrasound propagation direction.
The proposed method was trained with 102 labeled data set and tested to 70 ultrasound images.
We achieved a success rate of 92.31\% for HC and BPD estimations, and an accuracy of 87.14\% for the plane acceptance check.
\end{abstract}

}

\maketitle

\IEEEdisplaynontitleabstractindextext

\IEEEpeerreviewmaketitle

\section{Introduction}\label{sec:introduction}
Ultrasound-based fetal measurements have been the most widely used technique in the field of obstetrics, owing to its non-invasive real-time monitoring nature. For estimation of gestational age \cite{Chalana1996} and diagnosis of fetal growth disorders and cerebral anomalies \cite{ISOUG2007}, clinicians have used fetal biometric measurements such as biparietal diameter (BPD), head circumference (HC), and abdominal circumference (AC).
So far, fetal biometric measurements with ultrasound are performed manually and the accuracy of the measurement depends on two factors. The clinician should 1) acquire a proper plane and 2) the caliper must be properly positioned. However, these factors are operator-dependent and time-consuming involving multiple keystrokes and probe motions.
Hence, to improve the workflows and reduce user variability in data acquisition, there has been an increasing demand for automatic estimation of fetal biometric parameters \cite{Kurjak2017,Espinoza2013}.

However, developing the fully automated method is a very challenging problem, owing to inherent nature of ultrasound; patient specific and operator-dependent images affected by signal dropouts, artifacts, missing boundaries, attenuation, shadows, and speckling \cite{Rueda2014}. Moreover, these drawbacks can be obstacles to automatic navigation to a desired standard plane, which is essential for accurate biometric measurements.

To deal with these problems, there have been several researches on automatic estimation of fetal biometry.
In most methods, image intensity-based or gradient-based methods have been preferred to extract the boundaries of target anatomies \cite{Pathak2000, Lu2005, Yu2008, Ponomarev2012, Stebbing2012, Foi2014}.
To deal with the noisy ultrasound images, these methods use several approaches, including K-mean classifier, boundary fragment model, morphological operators, and minimizing the cost function.
However, the existence of surrounding tissues with intensities that are similar to those of the skull reduced the robustness of the head boundary detection.

Recently, with great successes in object recognition, machine learning methods have attracted much attention and was also applied in fetal biometry to analyze high-level features from ultrasound image data.
Carneiro {\it et al.} \cite{Carneiro2008} used probabilistic boosting tree to estimate the fetal biometric prarameters by classifying segment structures in ultrasound images. Although this approach showed some notable results, it requires complex, well-annotated data to train the tree.
Li {\it et al.} \cite{Li2017} utilized a random forest classifier to licalize the fetal head, then used phase symmetry and Ellifit to fit the HC ellipse for measurement. However, this method requires the prior knoledge about the gestational age and ultrasound scanning depth.
Wu {\it et al.} \cite{Wu2017} assessed image quality of fetal ultrasound image using two deep convolutional neural network (CNN) moedels. However, this approach implements only a part of an entire measurement process and need to be integrated for full automation of the measurement process.

This paper deals with the above challenging problem with focus on automated measurements of BPD and HC. We propose a deep-learning-based method reflecting the measurement process of clinicians to tackle the automated problem of the fetal biometry measurement with a high degree of accuracy and reliability. In this method, the input is ultrasound images containing the head of a fetus in the second or third trimester. The output is an image of an acceptable plane for HC and BPD and their measurements.
The learning algorithm must take into account the following requirements \cite{Hadlock1982, Shepard1982}:
1) BPD and HC are measured on the transthalamic plane, which shows the ``box-like" cavum septum pellucidum (CSP) and the ``V-shaped" ambient cistern (AC) but should not show the entire cerebellum (Cbll).
2) Calipers of BPD are placed on the outer edge of the near calvarial wall to inner edge of the far calvarial wall, and the line joining them is orthogonal to the central axis of the head.
3) HC is estimated by calculating the boundary of an ellipse drawn around the outside of the calvarium.

The most challenging part is robust detection of accurate head boundary from the ultrasound image.
We need to deal with large intra-class variations and artifacts caused by abutting structures and shadowing that increases with advancing gestation.
To overcome these difficulties, we developed a hierarchical perception model by exploiting prior knowledge of the fetal head anatomy.
We transform the images to differentiate tissue image patterns with respect to the ultrasound propagation direction.
For each transformed image, the head boundary pixels are detected using U-Net. To remove possible errors in the identification of the head boundary, we adopt bounding-box regression based on the two-stage procedure \cite{Mask R-CNN}. The ultrasound image is cropped to the fetal head region by fitting an ellipse to the detected head boundary pixels. We select the most acceptable plane among the cropped images by examining its internal anatomical structures. Finally, BPD and HC are measured from the selected plane.

The proposed method was trained with 102 labeled data set and tested on 70 ultrasound images. The experimental result shows that our method achieves good performance in ellipse fitting to head boundary, and shows potential to produce a high success rate of the plane acceptance check.

\begin{figure*}[ht]
	\centering
	\includegraphics[width=.9\linewidth]{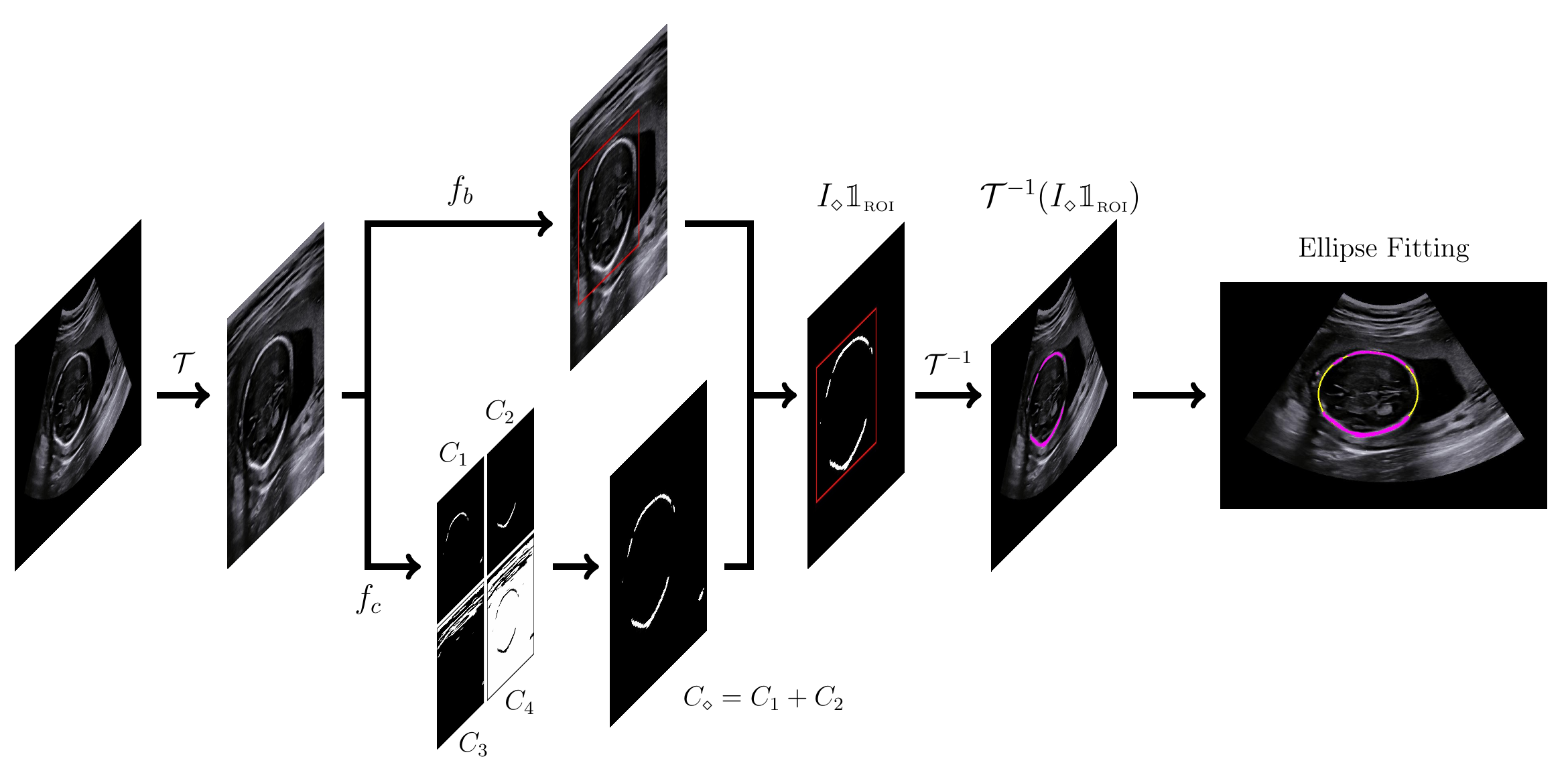}
	\caption{The overall process of the proposed method for head boundary detection.}
	\label{fig:framework}
\end{figure*}

\section{Method}\label{sec:method}
The goal is to find a function $f:\x \mapsto\y$, which maps from an ultrasound image $\x$ containing the head of a fetus to the output $\y$ consisting of HC and BPD measurements, and the acceptability score.
The proposed method is a hierarchical process that mimics a clinician's procedure for the fetal head biometric measurement. The framework of our method can be divided into 3 parts: (1) BPD and HC measurements, (2) Plane acceptance check, and (3) Measurement refinement.

\subsection{Head boundary detection}\label{subsec:Head boundary detection}
\begin{figure}[t]
	\centering
	\begin{tabular}{c c}
		\centering
		\includegraphics[width=0.45\linewidth]{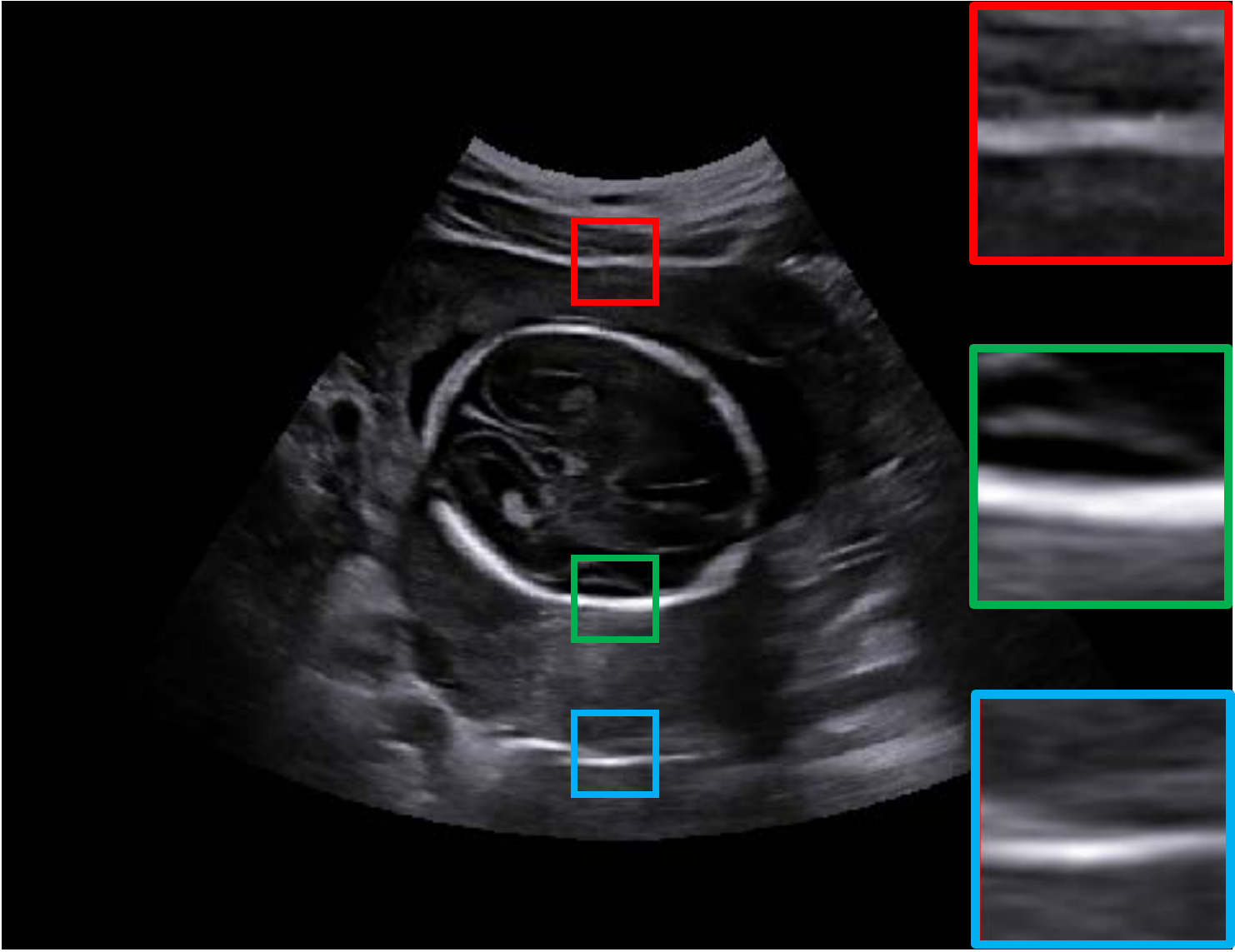} &
		\includegraphics[width=0.45\linewidth]{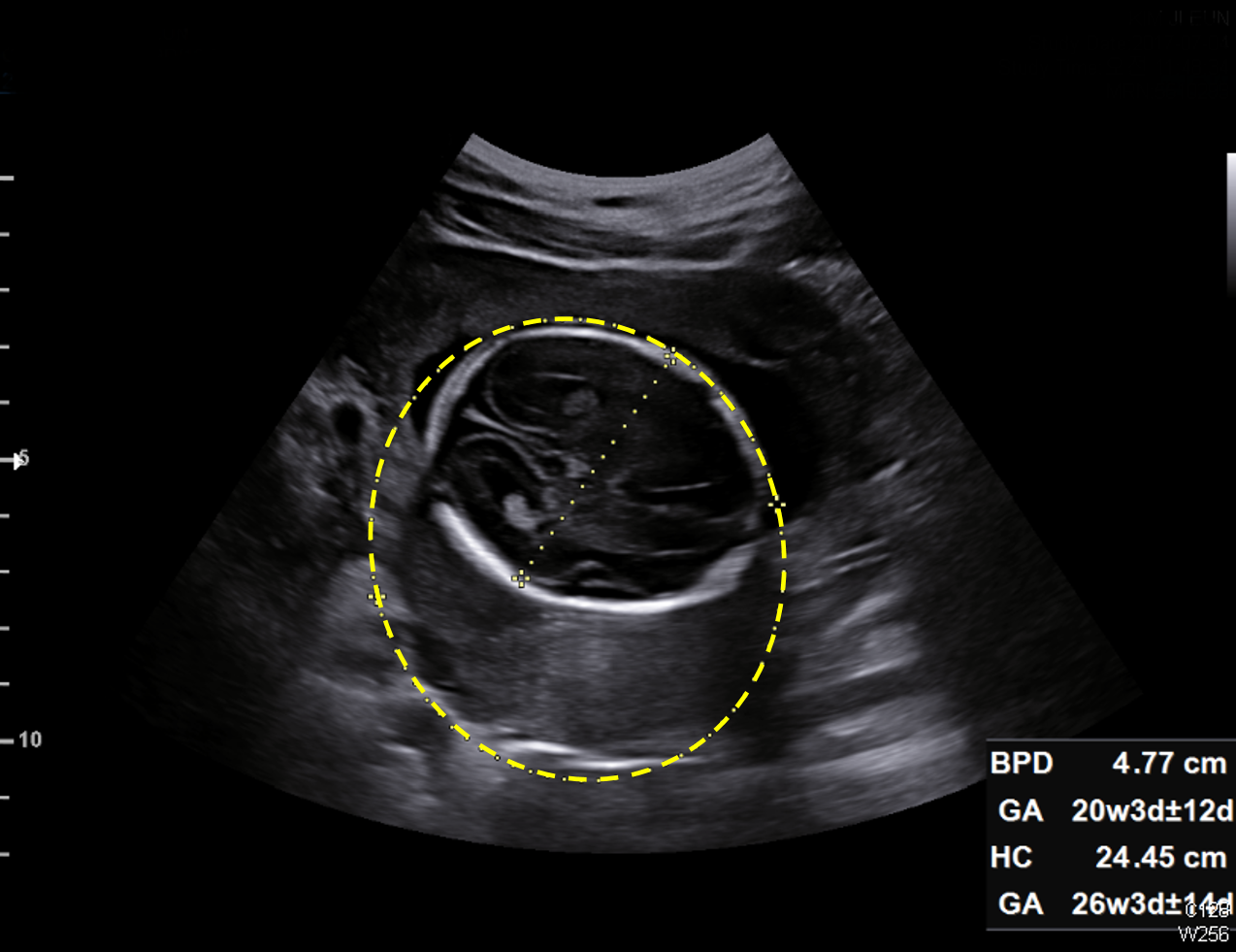} \\
		(a) & (b)
		
	\end{tabular}
	\caption{The difficulties of discrimination between head bounday and non-boundary pixels.
		(a) Each patch in ultrasound image has similar local patterns with distinct non-local structures,
		(b) a result of incorrect automated HC caliper placement by a commercially available semiautomated program due to wrong detection of head boundary.}
	\label{fig:commercial}
\end{figure}

The fetal head biometry detection uses an ellipse fitting method for detecting the head contour of the fetus in a propoer plane, which aims to place an ellipse around the outside of the skull bone echoes \cite{ISOUG2007}.
And proper plane is recommended by the guideline which is the cross-sectional view of the fetal head at the level of the thalami with symmetrical appearance of both hemispheres broken midline echo continuity by the CSP and without cerebellar visulization \cite{Salomon2011}.
For ellipse fitting, it is necessary to detect a suitable amount of head boundary pixels \cite{Xu1990, MC1998, Bennett1999, Ellifit}. Unfortunately, it is difficult to discriminate between head boundary and non-boundary pixels. As shown in Fig. \ref{fig:commercial}(a), it is difficult to know which one out of three patches contains the head boundary, because all three patches appear to have patterns associated with the head boundary. Fig. \ref{fig:commercial}(b), a result of automated HC measurement by a commercially available semiautomated program, shows an example of the incorrect automated caliper placement
caused by falsely perceiving the placenta boundary (blue box in Fig. \ref{fig:commercial}(a)) as head boundary.

To overcome this problem, we take advantage of knowledge of the ultrasound propagation direction and the image feature of the maternal tissues near the probe.
\subsubsection{Image transformation with respect to the ultrasound propagation direction}\label{subsubsec:Image transformation}
\begin{figure}[t]
	\centering
	\begin{tabular}{c c}
		\centering
		\includegraphics[height=4.5cm]{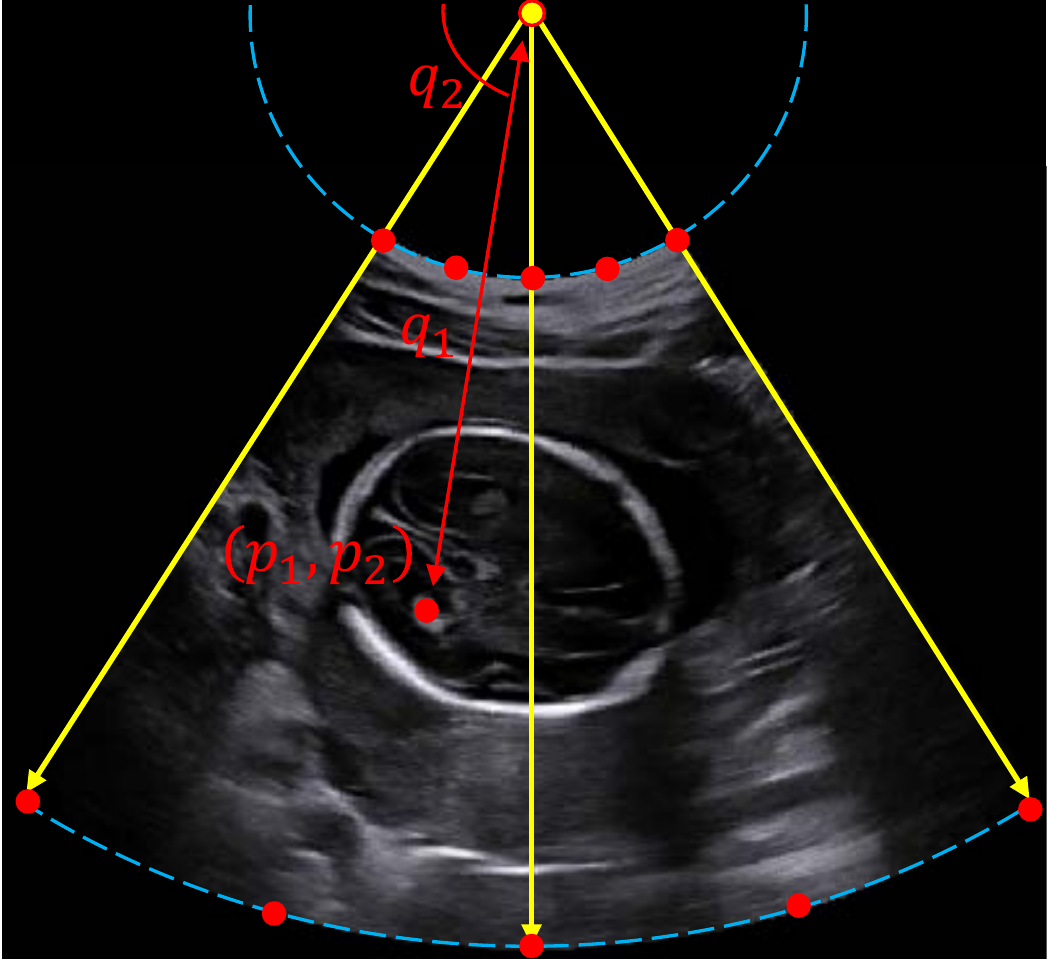} &
		\includegraphics[height=4.5cm]{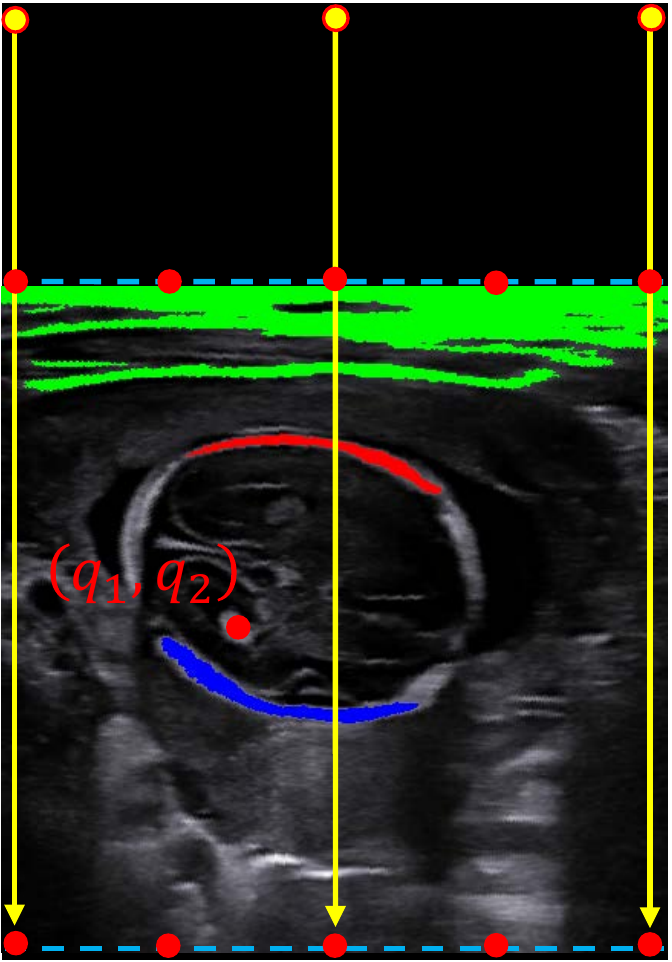} \\
		(a) & (b)
	\end{tabular}
	\caption{
		The ultrasound images with propagation direction.
		(a) Input image $\x$.
		(b) The transformed image $\mathcal{T}\x$.
		By taking the image transformation, the tissue image pattern (green at (b)) was effectively differentiated from the head boundary pattern.
	}
	\label{fig:polar_transform}
\end{figure}
For the maternal tissues in the near field, the image pattern is directed perpendicularly to the ultrasound propagation direction, as shown in Fig. \ref{fig:polar_transform}(a). Hence, if we transform the image from the Cartesian to polar coordinate system such that the ultrasound propagation direction is aligned to $y$-direction (vertical direction), then the image pattern of maternal tissue at near depth is directed to $x$-direction. Fig. \ref{fig:polar_transform}(a) shows an ultrasound image $\x$, where $\x(p_1,p_2)$ represents the gray-scaled values at position $(p_1,p_2)$ in rectangular coordinates. Fig. \ref{fig:polar_transform}(b) shows the image $\mathcal{T}\x$, corresponding to $\x$, transformed into polar coordinates. Note that $\mathcal{T}\x(q_1,q_2)=\x(p_1,p_2)$ if $q_1=\sqrt{p_1^2+p_2^2}$ and $q_2=\tan^{-1}{(p_2/p_1)}$.

Taking advantage of the transformed image $\mathcal{T}\x$, it is easy to extract the maternal tissue near the probe, because this tissue has distinct patterns (with horizontal direction) from the head boundary (See Fig. \ref{fig:polar_transform}(b)). By removing the maternal tissue from $\mathcal{T}\x$, it is possible to provide the robust detection of the head boundary. In terms of the ultrasound propagation direction, the first concave arc (the red arc at Fig. \ref{fig:polar_transform}(b)) and first convex arc (the blue arc at Fig. \ref{fig:polar_transform}(b)) structures indicate a portion of the head boundary.

Now, we are ready to detect the head boundary pixels from the transformed image $\mathcal{T}\x$. The goal is to learn a function $f_h:\mathcal{T}\x\mapsto\bP$ which maps from the transformed image $\mathcal{T}\x$ to the set of head boundary points $\bP$. Fig. \ref{fig:framework} illustrates the two-stage architecture of the networks utilized in our procedure.

\subsubsection{Multi-label pixel-wise classification map}\label{subsubsec:pixel-wise classification}
The first stage $f_c:\mathcal{T}\x \mapsto \CC$ adopts U-Net \cite{U-Net} to detect three different head boundary-like features; maternal tissues, upper head boundary, and lower head boundary. Here, the net $f_ c(\cdot,W_c)$ can be viewed as a function of weight parameters $W_c$. To be precise, the output $f_c(\mathcal{T}\x,W_c)$ divides the image into four binary images $\CC=(C_1,C_2,C_3,C_4)$, where $C_1$ is for the upper head boundary, $C_2$ is for the lower head boundary, $C_3$ is for the maternal tissue, and $C_4$ is for the remaining region, as shown in Fig \ref{fig:framework}.
Note that $C_k(q_1,q_2)=1$ if $\mathcal{T}\x(q_1,q_2)$ belongs to class $k$ at the spatial position $(q_1,q_2)$ in polar coordinate system and $C_k(q_1,q_2)=0$ otherwise.
	From the divided images $\CC$, we can simply obtain the head boundary $C_{\diamond}=C_1+C_2$ (see Fig \ref{fig:framework}).

The net $f_c(\cdot,W_c)$ is trained using a labeled training data $\{(\mathcal{T}\x^j, \I^j):j=1,\cdots,M\}$.
The weight parameter $W_c$ is determined by solving
\begin{equation}
\argmin\limits_{W_c} \frac{1}{M}\sum\limits_{j=1}^M H_{\Omega}(f_c(\mathcal{T}\x^j,W_I), \I^j)
\end{equation}
where the average cross-entropy loss $H_{\Omega}$ is given by
\begin{equation} \label{eq:cost-seg}
H_{\Omega}(\bX,\bY) = - \frac{1}{|\Omega|}\sum\limits_{(q_1,q_2)\in\Omega}\sum\limits_{k=1}^4\bX_k(q_1,q_2)\log{(\bY_k(q_1,q_2))},
\end{equation}
where $\Omega$ is a pixel grid in polar coordinate system. The stochastic gradient descent method via backpropgation allows us to solve this problem \cite{LeCun1989, LeCun2015}.

\begin{table}
	\centering
	\caption{U-Net architecture of $f_c$ for pixel-wise classification}
	\label{table:architecture-seg}
	\begin{tabular}{c| c c c c}
		\hline\hline
		\multirow{2}{*}{Input $\mathcal{T}\x$} & \multicolumn{4}{c}{Transformed image} \\ \cline{2-5}
		& \multicolumn{4}{c}{$224\times224\times1$} \\ \hline
		Layer & Type & Feature Maps & Filter size & Stride \\ \hline
		1 & conv$\times2$ & $224\times224\times64$ & $3\times3$ & 1 \\
		2 & maxpool & $112\times112\times64$ & $2\times2$ & 2 \\
		3 & conv$\times2$ & $112\times112\times128$ & $3\times3$ & 1 \\
		4 & maxpool & $56\times56\times128$ & $2\times2$ & 2 \\
		5 & conv$\times2$ & $56\times56\times256$ & $3\times3$ & 1 \\
		6 & maxpool & $28\times28\times256$ & $2\times2$ & 2 \\
		7 & conv$\times2$ & $28\times28\times512$ & $3\times3$ & 1 \\
		8 & maxpool & $14\times14\times512$ & $2\times2$ & 2 \\
		9 & conv & $14\times14\times1024$ & $3\times3$ & 1 \\
		10 & conv & $14\times14\times512$ & $3\times3$ & 1 \\
		11 & avg-unpool & $28\times28\times512$ & $2\times2$ & 2 \\
		12 & concat 7 & $28\times28\times1024$ & - & - \\
		13 & conv & $28\times28\times512$ & $3\times3$ & 1 \\
		14 & conv & $28\times28\times256$ & $3\times3$ & 1 \\
		15 & avg-unpool & $56\times56\times256$ & $2\times2$ & 2 \\
		16 & concat 5 & $56\times56\times512$ & - & - \\
		17 & conv & $56\times56\times256$ & $3\times3$ & 1 \\
		18 & conv & $56\times56\times128$ & $3\times3$ & 1 \\
		19 & avg-unpool & $112\times112\times128$ & $2\times2$ & 2 \\
		20 & concat 3 & $112\times112\times256$ & - & - \\
		21 & conv & $112\times112\times128$ & $3\times3$ & 1 \\
		22 & conv & $112\times112\times64$ & $3\times3$ & 1 \\
		23 & avg-unpool & $224\times224\times64$ & $2\times2$ & 2 \\
		24 & concat 1 & $224\times224\times128$ & - & - \\
		25 & conv$\times2$ & $224\times224\times64$ & $3\times3$ & 1 \\ \hline
		Output $\CC$ & conv & $224\times224\times4$ & $3\times3$ & 1 \\ \hline\hline
	\end{tabular}
\end{table}

\begin{figure}[t]
	\centering
	\includegraphics[width=0.9\linewidth]{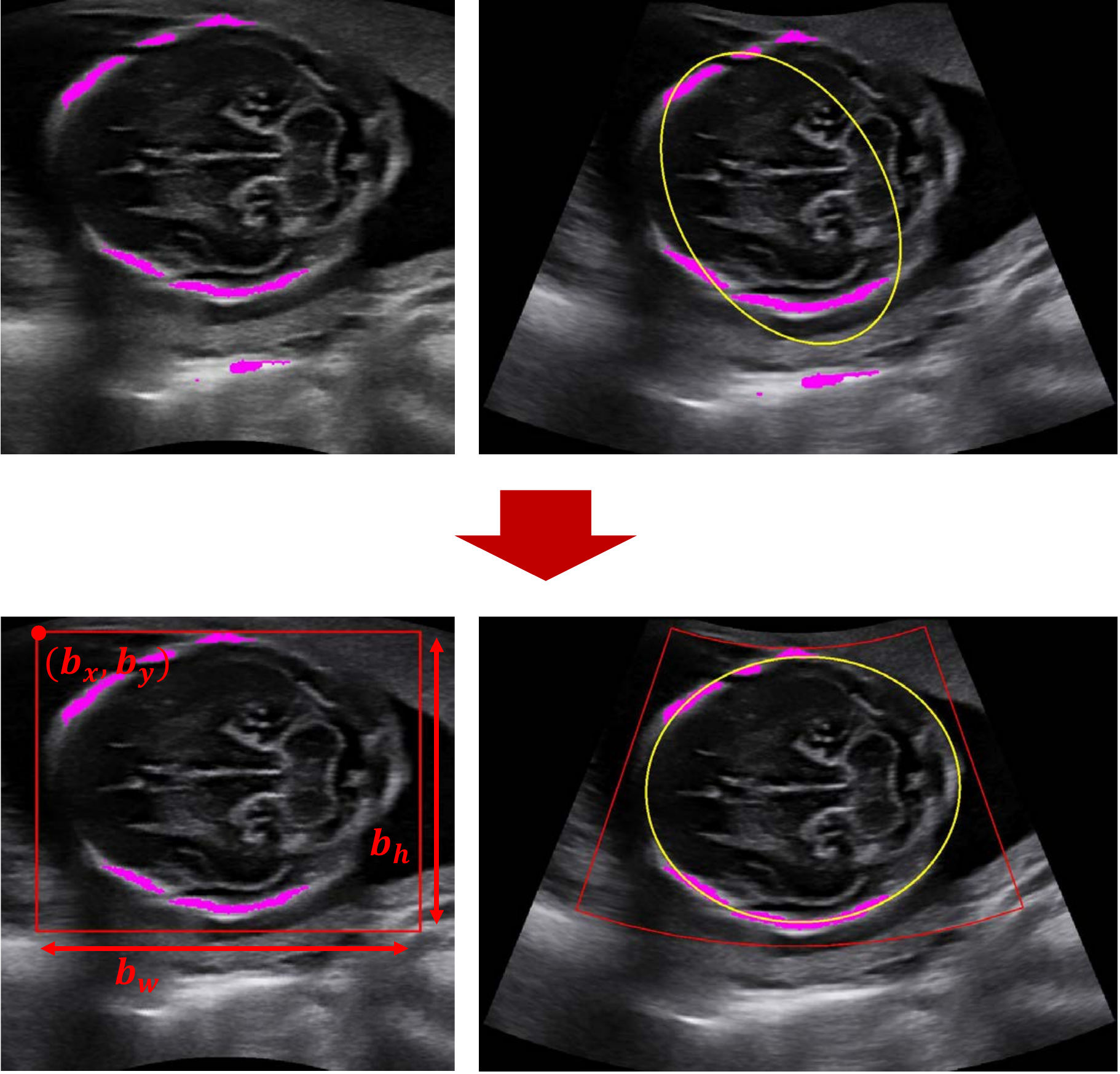}
	\caption{The role of bounding-box. The figure on top shows a wrong ellipse fitting due to the misclassified pixels. Bounding-box allows our method to robustly fit the ellipse by removing the non-boundary pixels.
	}
	\label{fig:bounding_box}
\end{figure}

\begin{remark}
{\it
Let us mention why we use the classification map $f_c$, dividing four classes, instead of segmenting head boundary directly. In our experiments, the most accurate way of detecting head boundary by U-Net is to classify each pixel into four different classes: maternal tissues having horizontal directional pattern, upper head boundary having concave arc pattern, lower head boundary having convex arc pattern, and the remaining. We tried to segment head boundary directly by classifying each pixel into two classes: head boundary and the remaining through the same U-Net, but we could not achieve a satisfactory learning with a reasonable accuracy.}
\end{remark}
\begin{remark}\label{remark:2}
{\it We should mention that the classification map $f_c$ may misclassify a few pixels. As shown in Fig. \ref{fig:bounding_box}, this may result in a wrong ellipse fitting when applying the method in \cite{Ellifit}.
It is desirable to filter out these misclassified pixels for a robust ellipse fitting.}
\end{remark}

To deal with the problem in {\it Remark \ref{remark:2}}, we use a bounding-box regression to remove wrongly classified pixels in the second stage.
\subsubsection{Bounding-box regression}\label{subsubsec:bounding-box}
The bounding-box regressor $f_b:\mathcal{T}\x\mapsto \b$ adopts VGG-Net \cite{VGG} and the net $f_b(\cdot,W_b)$ can be viewed as a function of weight parameters $W_b$. To be precies, the output $f_b(\mathcal{T}\x,W_b)$ takes the bounding-box coordinates $\b=(b_x,b_y,b_w,b_h)$, where $(b_x,b_y)$ is the top-left point and $(b_h,b_w)$ are the height and width, as shown in Fig. \ref{fig:bounding_box}.

This bounding-box coordinates filters out the misclassified pixels outside of it so that the head boundary pixels are restricted to $I_{\diamond}\, \mathbbm{1}_{\mbox{\tiny ROI}}$. Here, $\mathbbm{1}_{\mbox{\tiny ROI}}$ is the indicator function of $\mbox{\small ROI}=[b_x,b_x+b_w]\times[b_y,b_y+b_h]$. Finally, we obtain the set of head boundary points $\bP$ by transforming the head boundary image $I_{\diamond}\, \mathbbm{1}_{\mbox{\tiny ROI}}$ back into the coordinate system of ultrasound images, which is given by
\begin{equation}\label{eq:head boundary}
	\bP=\{(p_1,p_2): \left(\mathcal{T}^{-1}(I_{\diamond}\, \mathbbm{1}_{\mbox{\tiny ROI}})\right)(p_1,p_2)=1\}.
\end{equation}
For training the net $f_b(\cdot,W_b)$, we repeat a similar procedure as before. Let $\{(\mathcal{T}\x^j, \b^j):j=1,\cdots,M\}$ be a labeled training data.  Training the net is achieved by solving the minimization problem:
\begin{equation} \label{eq:cost-bbox}
\argmin\limits_{W_b} \frac{1}{M}\sum\limits_{j=1}^M \|f_b(\mathcal{T}\x^j,W_b) - \b^j \|^2
\end{equation}
where $\|\cdot\|$ denotes the Euclidean norm.

\begin{table}
	\centering
	\caption{CNN architecture of $f_b$ for bounding-box regression}
	\label{table:architecture-bbox}
	\begin{tabular}{c| c c c c}
		\hline\hline
		\multirow{2}{*}{Input $\mathcal{T}\x$} & \multicolumn{4}{c}{Transformed image} \\ \cline{2-5}
		& \multicolumn{4}{c}{$224\times224\times1$} \\ \hline
		Layer & Type & Feature Maps & Filter size & Stride \\ \hline
		1 & conv$\times3$ & $224\times224\times64$ & $3\times3$ & 1 \\
		2 & maxpool & $112\times112\times64$ & $2\times2$ & 2 \\
		3 & conv$\times3$ & $112\times112\times128$ & $3\times3$ & 1 \\
		4 & maxpool & $56\times56\times128$ & $2\times2$ & 2 \\
		5 & conv$\times3$ & $56\times56\times256$ & $3\times3$ & 1 \\
		6 & maxpool & $28\times28\times256$ & $2\times2$ & 2 \\
		7 & conv$\times3$ & $28\times28\times512$ & $3\times3$ & 1 \\
		8 & maxpool & $14\times14\times512$ & $2\times2$ & 2 \\
		9 & conv$\times3$ & $14\times14\times512$ & $3\times3$ & 1 \\
		10 & maxpool & $7\times7\times512$ & $2\times2$ & 2 \\
		11 & FC & 4096 & - & - \\
		12 & FC & 4096 & - & - \\
		13 & FC & 1000 & - & - \\ \hline
		Output $\b$ & FC & 4 & - & - \\ \hline\hline
	\end{tabular}
\end{table}

From the head boundary points $\bP$ in \eref{eq:head boundary}, it is easy to obtain HC and BPD measurements.
These measurements can be estimated by placing an ellipse around the outside of the skull. For reader's convenience, we will explain the ellipse fitting method \cite{Ellifit}, called Ellifit, in the following section.
\subsubsection{HC and BPD measurements}\label{subsec:HC and BPD measurements}
Ellifit is a least squares based geometric ellipse fitting method that obtains the five ellipse parameters $\Theta=(a,b,\theta_c,x_c,y_c)$, which provides the following ellipse representation
\begin{equation}
\alpha(x-x_c)^2+\beta(y-y_c)^2+\gamma(x-x_c)(y-y_c)=a^2b^2,
\end{equation}
where
\begin{align*}
\alpha &= a^2\sin^2\theta_c+b^2\cos^2\theta_c,\\
\beta &= a^2\cos^2\theta_c+b^2\sin^2\theta_c,\\
\gamma &= (a^2-b^2)\sin2\theta_c.
\end{align*}
The parameter $\Theta$ is obtained by solving
\begin{equation}\label{eq:Ellifit}
\argmin\limits_{\Theta}\sum\limits_{(p_1,p_2)\in\bP} \frac{\left|p_2-m(\Theta;p_1,p_2) p_1-c(\Theta;p_1,p_2) \right|^2}{1+m(\Theta;p_1,p_2)^2},
\end{equation}
where
\begin{equation*}
m(\Theta;p_1,p_2)= -\frac{2\alpha(p_1-x_c)+\gamma(p_2-y_c)}{2\beta(p_2-y_c)+\gamma(p_1-x_c)},
\end{equation*}
\begin{align*}
c(\Theta;p_1,p_2)=&\frac{1}{2\beta(p_2-y_c)+\gamma(p_1-x_c)} \nonumber \\
& \times \Bigl(2\beta(p_2-y_c)p_2+2\alpha(p_1-x_c)p_x \nonumber \\
& \quad ~  -\gamma(p_2x_c+p_1y_c-2p_1p_2)\Bigr).
\end{align*}
ElliFit splits above minimization problem into two operations such that the overall algorithm is non-iterative, numerically stable, and computationally inexpensive. For the details of the method, please refer to \cite{Ellifit}. Fig. \ref{fig:bounding_box} shows a result obtained by the ElliFit method (yellow contour).

Once $\Theta=(a,b,\theta_c,x_c,y_c)$ is determined by \eref{eq:Ellifit},  BPD and HC, respectively, are given by $\mbox{BPD}  = 2b$ and
\begin{align*}
\mbox{HC}  &= \pi \left[ 3(a+b)-\sqrt{(3a+b)(a+3b)}\right].
\end{align*}

\subsection{Plane acceptance check}\label{subsec:Plane acceptance check}
Based on the knowledge of the ellipse parameters $\Theta$ for each image, we developed a method of evaluating the suitability of the selected plane whether the plane is appropriate for measuring BPD and HC. This can be achieved by examining anatomic landmarks inside the fetal head region. However, this region is a relatively small area within the ultrasound image, and the image patterns outside of it may disturb the examination. We use the information of  $\Theta$ to reduce the search range for three feature points inside the head.

\begin{figure}[t]
	\centering
	\begin{tabular}{c c c}
		\includegraphics[height=1.95cm]{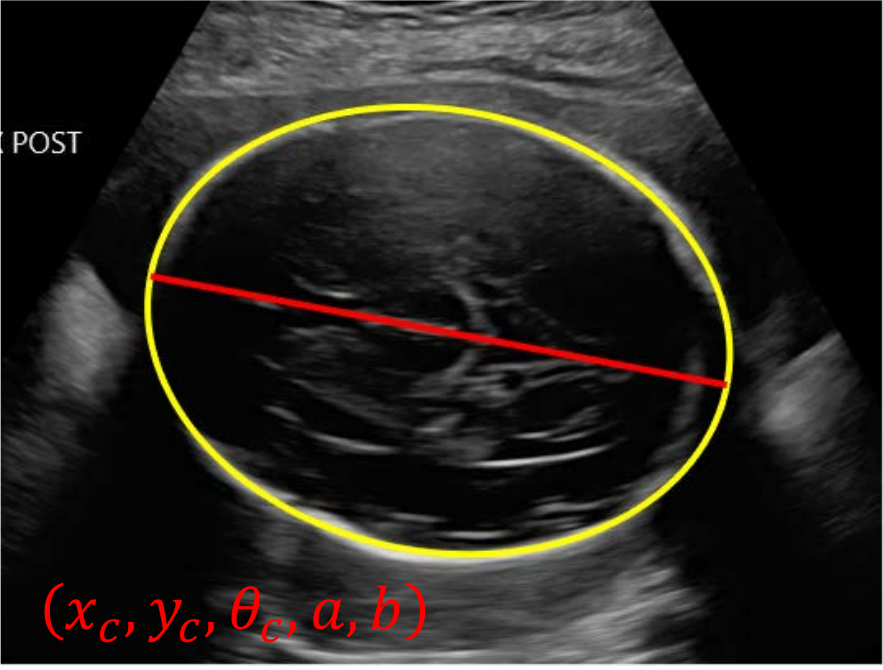} &
		\includegraphics[height=1.95cm]{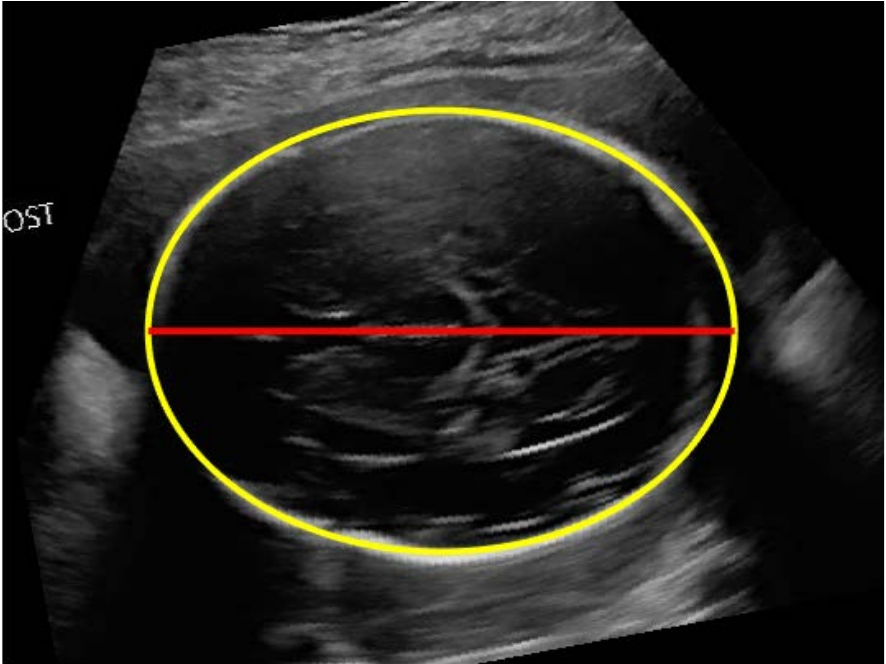} &
		\includegraphics[height=1.95cm]{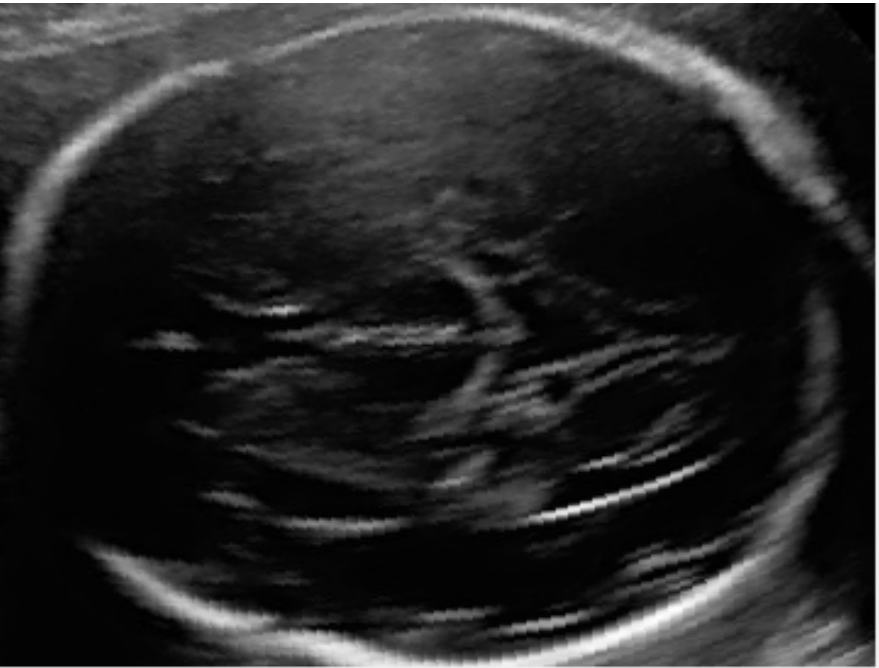} \\
		(a) & (b) & (c)
	\end{tabular}
	\caption{The process of the search space reduction to inside the head. (a) The ellipse (yellow contour) fitted to head boundary and its major axis (red line), (b) the rotated image to align the ellipse horizontally, and (c) the cropped image around the ellipse.}
	\label{fig:search space reduction}
\end{figure}

\subsubsection{Search space reduction}\label{subsubsec:Search space reduction}
\begin{figure*}[ht]
	\centering
	\addtolength{\subfigcapskip}{0.05in}
	\begin{tabular}{c|c|ccc}
		Standard plane & Accepted case & \multicolumn{3}{c}{Rejected case} \\ \cline{2-5}
		\subfigure{\includegraphics[height=0.15\textwidth]{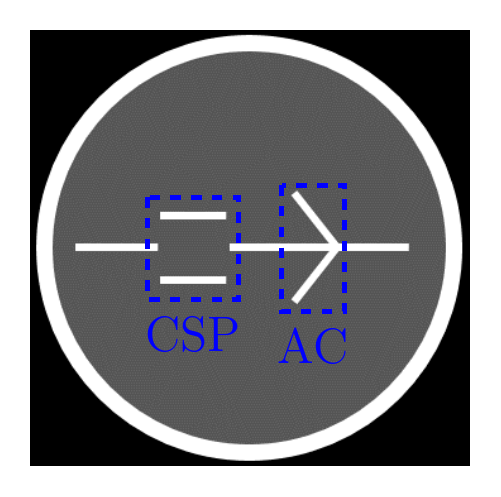}}
		&\subfigure{\includegraphics[height=0.15\textwidth]{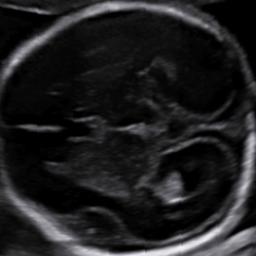}}
		& \subfigure{\includegraphics[height=0.15\textwidth]{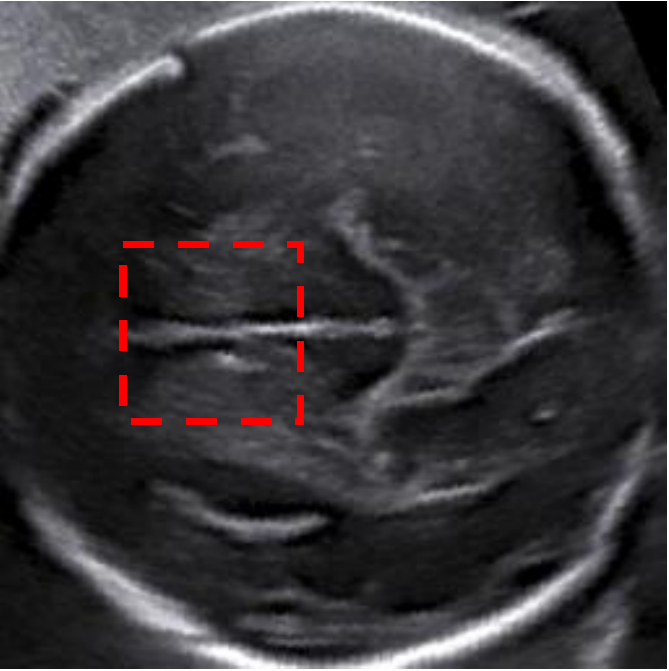}}
		& \subfigure{\includegraphics[height=0.15\textwidth]{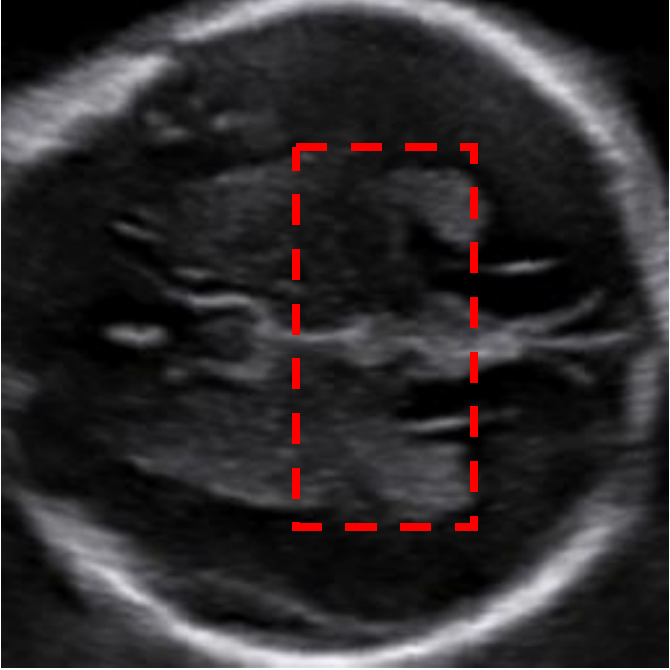}}
		& \subfigure{\includegraphics[height=0.15\textwidth]{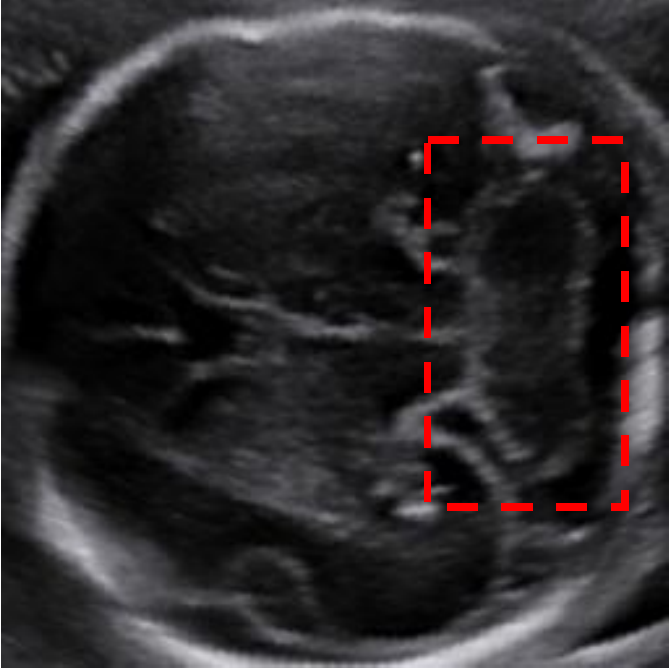}} \\ \hline
		CSP & 1 & 0 & 1 & 1 \\ \hline
		AC & 1 & 1 & 0 & 1 \\ \hline
		Cbll & 1 & 1 & 1 & 0 \\ \hline
	\end{tabular}
	\caption{Fetal head ultrasound images and anatomical structures. In a standard plane, the ``box-like" cavum septum pellucidum (CSP) and the ``V-shaped" ambient cistern (AC) must be demonstrated and cerebellum (Cbll) should not be visible. The red dotted boxes show the deduction factor of the acceptability score.}
	\label{fig:AC}
\end{figure*}

We use prior knowledge of the geometric placement of the anatomic landmarks. In order to normalize the geometric placements, we rotate the image by $\theta_c$ in \eref{eq:Ellifit}, so that  the image is aligned in terms of the direction of fetus' midline. We crop this rotated image around the ellipse to obtain $\x_{\Theta}$, as shown in Fig. \ref{fig:search space reduction}. From this cropped image, it is much easier to examine the landmarks within the head region that the landmarks are positioned at almost same area over the different images, as shown in Fig. \ref{fig:AC}.

\subsubsection{Acceptability scoring}\label{subsubsec:Acceptability scoring}
Three components were assessed for image scoring: (1) cavum septum pellucidum, (2) ambient cistern, and (3)  cerebellum. Normally visible structures were assigned a score of 1 each, otherwise a score of 0, see Table \ref{table:scoring}. The cropped image $\x_{\Theta}$ will be regraded as a standard plane if it gets 3 points.
We use the CNN $f_s:\x_{\Theta}\mapsto \s=(s_1,s_2,s_3)$ to learn this scoring process, where $s_1$ is the score of CSP, $s_2$ is the score of AS, and $s_3$ is the score of Cbll. The net $f_s(\cdot,W_s)$ adopts the VGG-Net and trained usisng a labeled training data $\{(\x_{\Theta}^j, \s^j):j=1,\cdots,M \}$, which is achieved by solving
\begin{equation}\label{eq:cost-acc}
	\argmin_{W_s} \frac{1}{M} \sum_{j=1}^M \|f_s(\x^j_{\Theta},W_s)-\s^j\|^2.
\end{equation}

\begin{table}[ht]
	\caption{Quality Assessment criteria for fetal head ultrasound images. Normally visible structures are assigned a score of 1 each, otherwise a score of 0.}
	\centering
	\label{table:scoring}
	\begin{tabular}{c|c | c}
		\hline\hline
		Component & Criteria & Score \\ \hline
		\makecell{Cavum septum \\ pellucidum (CSP)} & \makecell{Two parallel echogenic lines \\ between falx and the thalami.} & 1 \\ \hline
		Ambient cistern (AC) & \makecell{V-shaped echogenic line converging \\ behind the paired thalami.} & 1 \\ \hline
		Cerebellum (Cbll) & \makecell{Cerebellum is either not visible \\ or vermis is partially visible.} & 1 \\ \hline\hline
	\end{tabular}
\end{table}
\begin{table}
	\centering
	\caption{CNN architecture of $f_s$ for acceptability scoring}
	\label{table:architecture-acc}
	\begin{tabular}{c| c c c c}
		\hline\hline
		\multirow{2}{*}{Input $\x_{\Theta}$} & \multicolumn{4}{c}{Cropped image} \\ \cline{2-5}
		& \multicolumn{4}{c}{$256\times256\times1$} \\ \hline
		Layer & Type & Feature Maps & Filter size & Stride \\ \hline
		1 & conv$\times3$ & $256\times256\times64$ & $3\times3$ & 1 \\
		2 & maxpool & $128\times128\times64$ & $2\times2$ & 2 \\
		3 & conv$\times3$ & $128\times128\times128$ & $3\times3$ & 1 \\
		4 & maxpool & $64\times64\times128$ & $2\times2$ & 2 \\
		5 & conv$\times3$ & $64\times64\times256$ & $3\times3$ & 1 \\
		6 & maxpool & $32\times32\times256$ & $2\times2$ & 2 \\
		7 & conv$\times3$ & $32\times32\times512$ & $3\times3$ & 1 \\
		8 & maxpool & $16\times16\times512$ & $2\times2$ & 2 \\
		9 & conv$\times3$ & $16\times16\times512$ & $3\times3$ & 1 \\
		10 & maxpool & $8\times8\times512$ & $2\times2$ & 2 \\
		11 & FC & 4096 & - & - \\
		12 & FC & 4096 & - & - \\
		13 & FC & 1000 & - & - \\ \hline
		Output $\s$ & FC & 3 & - & - \\ \hline\hline
	\end{tabular}
\end{table}

\begin{figure*}[ht]
	\centering
	\includegraphics[width=.95\linewidth]{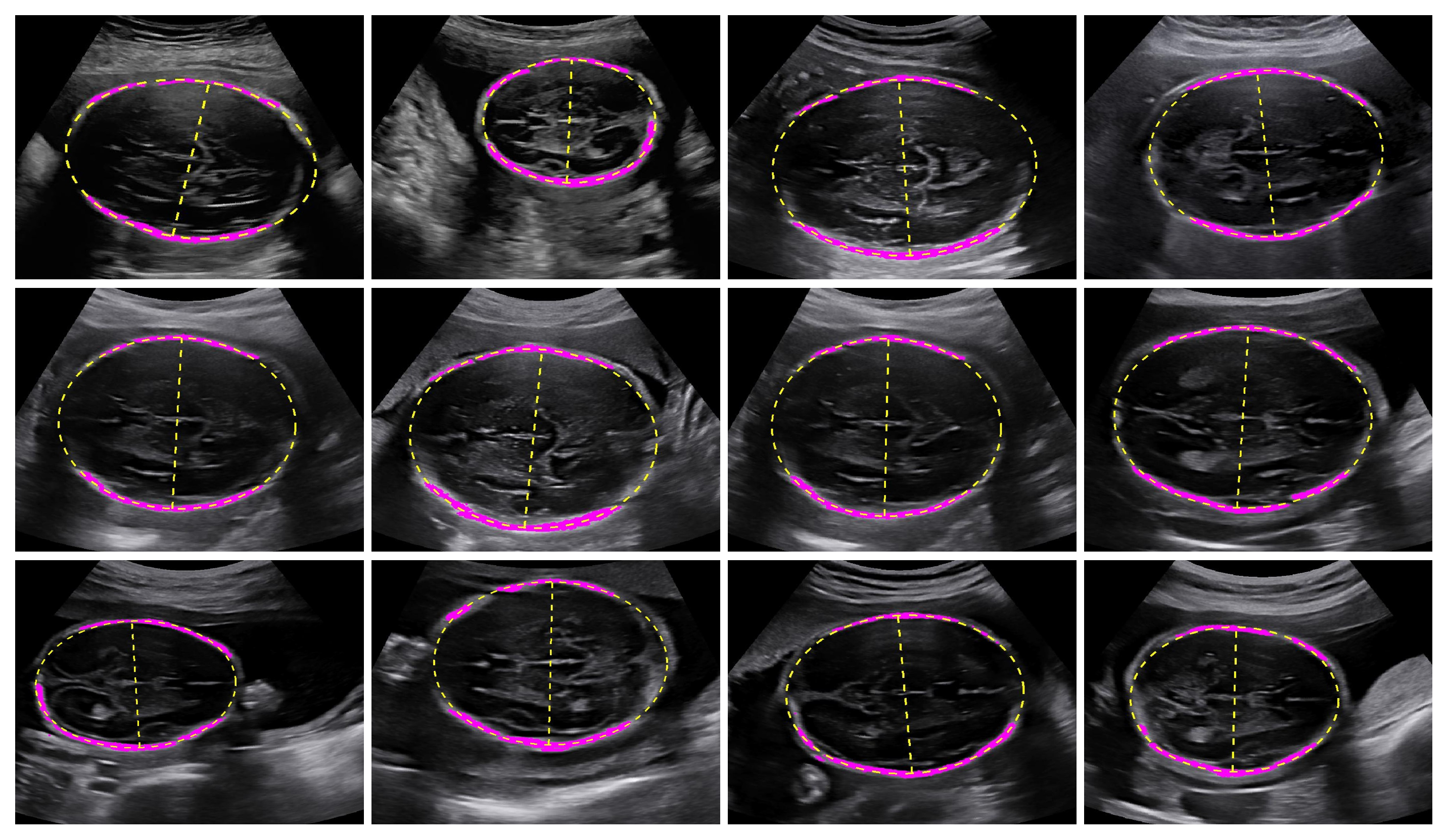}
	\caption{The results for BPD and HC measurements accepted by the experts. The magenta region denotes the detected head boundary points. Yellow dotted line and ellipse denotes the caliper placements of the proposed method for BPD and HC, respectively.}
	\label{fig:HC&BPD-good}
\end{figure*}

\section{Experiments and Results}\label{sec:results}

\subsection{Experimental Setting}\label{subsec:results-setting}
For training and evaluation, fetal head ultrasound images were provided by the department of Obstetrics and Gynecology, Yonsei university college of medicine, Seoul, Korea (IRB no.: 4-2017-0049). The ultrasound images were obtained by experts with an WS80A (SAMSUNG Medison, Seoul, Korea) ultrasound machine using a 2-6-MHz transabdominal transducer CA1-7A.

The provided images include 102 ultrasound images for training of each neural networks, and 70 ultrasound images for evaluation of fetal head biometry and plane acceptance check. The same set of images for evaluation were provided to ultrsound expert1 (J.-Y Kwon) and expert2 (Y. J. Park) for manual assessment blinded to each other's result. Training and test datasets include 19 and 13 of 2D axial images of the true transthalamic plane images. There are 46, 65, and 28 positive samples of CSP, AS, and Cbll for training dataset, and 33, 38, and 20 positive samples for test dataset.

The cost functions \eref{eq:cost-seg}, \eref{eq:cost-bbox} and \eref{eq:cost-acc} were minimized using the RMSPropOptimizer \cite{RMSProp} with a learning rate of $0.0001$, weight decay of $0.9$, and mini-batch size of 32 at each epoch. We used a trained networks with 500 training epochs. Training was implemented by Tensorflow r1.8 \cite{Tensorflow} on a CPU (Intel(R) Core(TM) i7-6850K, 3.60GHz) and a four GPU (NVIDIA GTX-1080, 8GB) system running Ubuntu 16.04.4 LTS. The networks required approximately 10, 8 and 8 hours for training, respectively. Our framework which consists of the U-Net, CNNs and the Ellifit was implemented with MATLAB (R2017b) and Python (3.5.3).

\subsection{Training data acquisition}\label{subsec:results-data}
For the training data, we transformed the images to $\mathcal{T}\x$ and labeled the pixel-wise classification maps and bounding box coordinates from the provided ultrasound images.
We received the ultrasound images in the size range of $509\times757$ to $800\times1088$.
After taking image transformation, the size of images were reduced to $224\times224$.
To obtain the ground-truth pixel-wise classification map $\CC$, we applied global thresholding of various threshold values to the transformed images. Using the flood fill algorithm to each binarized image, we obtain three head boundary-like features ($C_1, C_2, C_3$) and remaining region $C_4$.
To avoid the time consuming process of labeling, we developed a data collecting program using MATLAB, which can collect the labeled data semi-automatically.

In order to reduce overfitting and make the network more robust to varying object's position and size, we augmented the training data by cropping and flipping. For each transformed image, we cropped 50 square patches by changing its position and size, then scaled the patches up to $224\times224$ pixels. The labels corresponding to each patch were generated in the same way.

We also augmented the training data for the plane acceptance check.
From an ultrasound image and a detected ellipse parameters, we generated 25 cropped images by adding small random noise on the ellipse parameters, then scaled the images up to $256\times256$ pixels.

\begin{figure*}[ht]
	\centering
	\includegraphics[width=\linewidth]{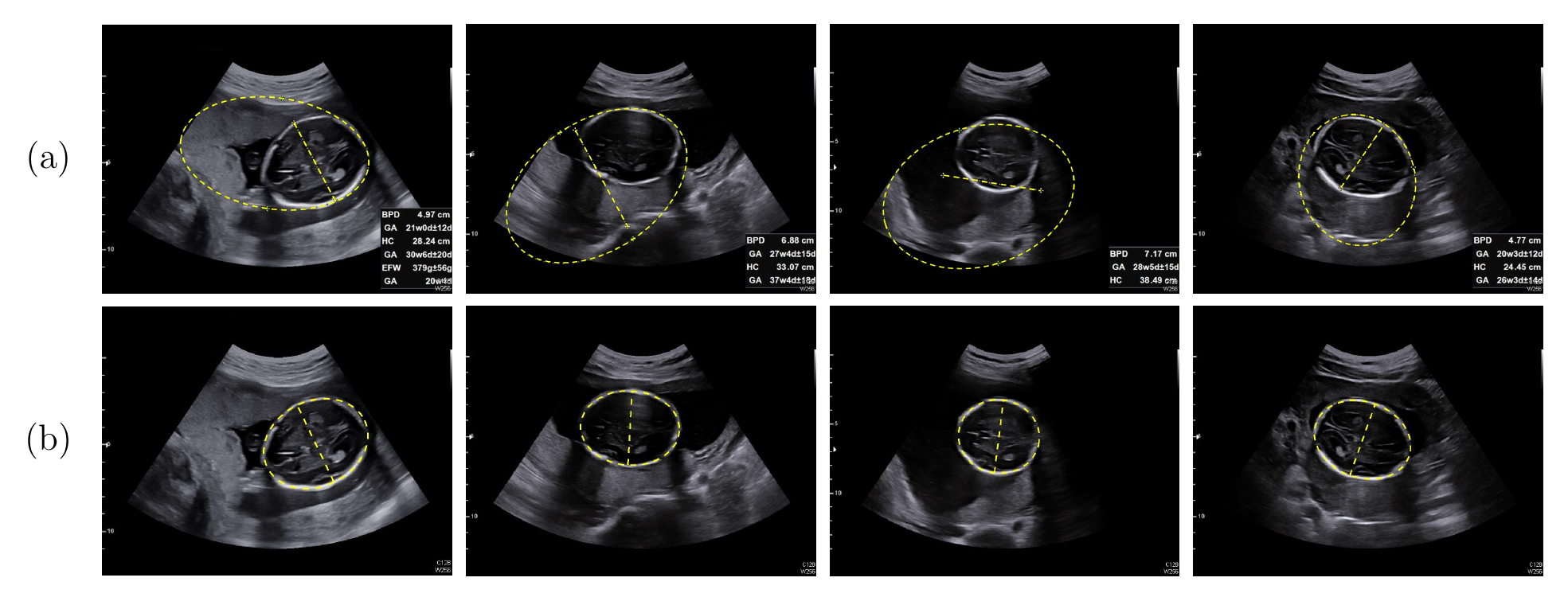}
	\caption{Comparison between the proposed method and a commercially available semiautomated progrm: (a) the incorrect cases of the automated caliper placements from a commercially available semiautomated program and (b) the result of the proposed method for HC and BPD measurements.}
	\label{fig:comparison-commercial}
\end{figure*}

\begin{figure}[t]
	\centering
	\includegraphics[width=.8\linewidth]{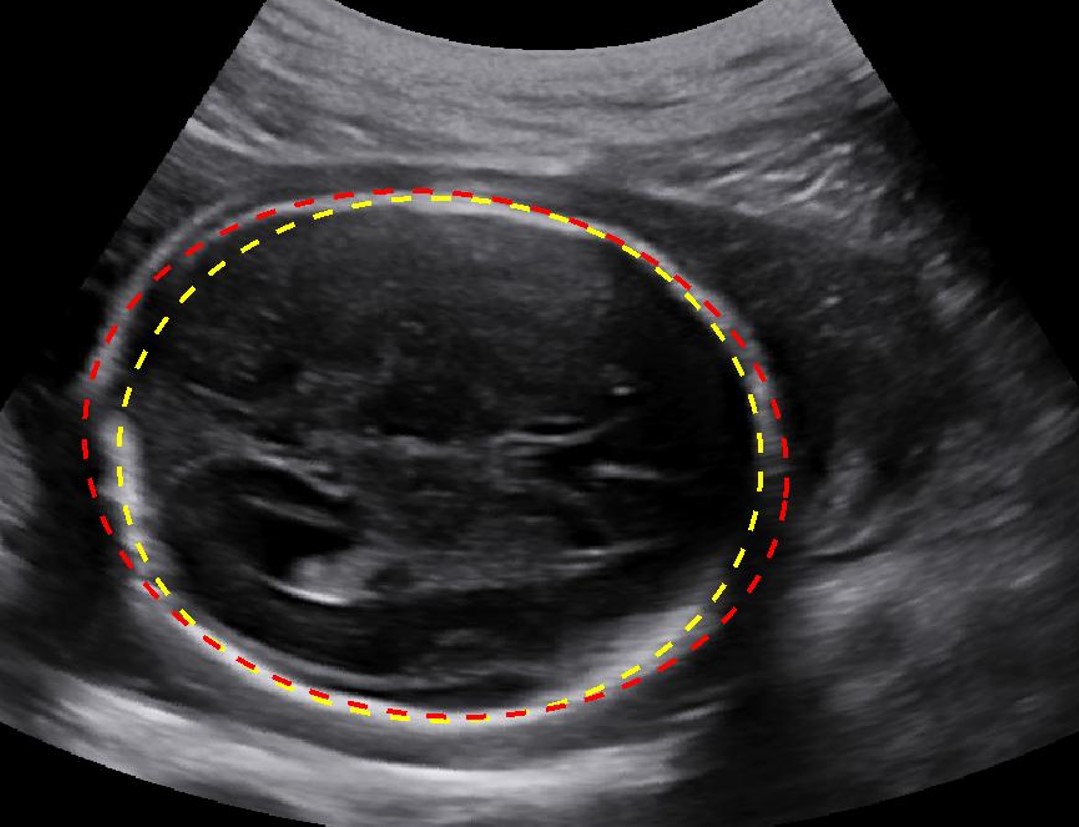}
	\caption{A result of HC measurement rejected by an expert. Yellow dotted ellipse denotes our result and red dotted ellipse denotes experts' caliper placement.}
	\label{fig:HC&BPD-bad}
\end{figure}

\subsection{HC and BPD Measurements}\label{subsec:results-HC&BPD}
For the assessment of HC and BPD measurements, the ultrasound images captured by the experts were used to estimate the measurements. A proper axial image was defined as the cross-sectional view of the fetal head at the level of the thalami with symmetrical appearance of both hemispheres and with continuous midline echo (falx cerebri) broken in middle by the CSP and thalamus \cite{Salomon2011}. And we defined well-visualized CSP as parallel lines or box-like structure located at anterior to thalami.
From the estimated measurements, the calipers for HC and BPD were placed on the ultrasound images and its goodness was assessed by the experts.
Caliper placement for HC annotated as `correct' by the experts if ellipse was placed around external border of the cranium echoes. Caliper placement for BPD was classified as `correct' if both calipers are placed from outer edge to inner edge, at the widest part of the skull, with perpendicular angle to the midline falx \cite{Salomon2011}.
We achieved a success rate of 92.31\% for HC and BPD estimations.
Fig. \ref{fig:HC&BPD-good} shows the results of the head boundary detection and caliper placements accepted by the experts.

In addition, we tested the proposed method on the cases where the automated caliper placement by a commercially available semiautomated program was annotated as `incorrect' by the experts.
As shown in Fig. \ref{fig:comparison-commercial}(a), the automated calipers from a commercially available semiautomated program were completely misplaced due to the wrong detection of head boundary. The proposed method shows even better results for such cases.

Fig. \ref{fig:HC&BPD-bad} shows a result of HC measurement rejected by an expert.
The caliper should have been placed around the outer margin of skull echo.
Although this result was not accepted, unlike the heavy outliers in Fig. \ref{fig:comparison-commercial}(a), it was properly fitted within certain margins from the head boundary.

\subsection{Plane acceptance check}\label{subsec:results-acceptance}
To measure the performance of the plane acceptance check, we compared our result of acceptability scoring with the annotation from two experts.
The proposed method was quantitatively evaluated using three assessment metrics of specificity (true negative rate), sensitivity (true positive rate) and accuracy.

Table \ref{table:AC} shows the comparison of acceptance check between the proposed method and experts' annotation in terms of CSP, AC, Cbll, and overall acceptance. The agreement between the proposed method and experts was 87.14\% while the agreement between two experts was 100\%.

\begin{table*}
	\caption{Comparison of acceptance check between the proposed method and experts' annotations}
	\label{table:AC}
	\centering
	\begin{tabular}{c | c c c c | c c c c | c c c c}
		\hline\hline
		\multirow{2}{*}{} & \multicolumn{4}{c|}{Specificity (\%)} & \multicolumn{4}{c|}{Sensitivity (\%)} & \multicolumn{4}{c}{Accuracy (\%)} \\ \cline{2-13}
		& CSP  & AC  & Cbll  & Acceptance & CSP & AC  & Cbll & Acceptance & CSP  & AC & Cbll & Acceptance \\ \hline
		Expert1 / Expert2 & - & - & - & - & - & - & - & - & 92.9 & 98.6 & 100 & 100 \\
		Proposed / Expert1 & 72.7 &  93.8 & 100 &  94.7 & 78.4 & 89.5 & 100 & 53.9 & 75.7 & 91.4 & 100 &  87.1\\
		Proposed / Expert2 & 68.4 & 96.8 & 100 & 94.7 & 81.3 & 89.7 & 100 & 53.9 & 74.3 & 92.9 & 100 & 87.1\\
		\hline\hline
	\end{tabular}
\end{table*}

\section{Discussion and conclusion}\label{sec:discussion}
This paper proposes a deep-learning-based method for automatic evaluation of fetal head biometry from ultrasound images. We differentiated between head boundary pixels and non-boundary pixels using image transformation, so that it is possible to segment more efficiently through U-net. Also, we adopt bounding-box regression to remove wrongly classified pixels. In our experiments, only ultrasound images were provided without probe geometry which are available when the method is implemented into a ultrasound system. Due to the absence of the information, we obtained the image transformation $\mathcal{T}$ by the following process. First, the origin is selected by choosing three points on top of the ultrasound image, then fitting a circle that passes through chosen points. Next, we randomly select four points on each side of the ultrasound image to determine the range of radius and angle.

The experimental results show that our method achieves good performance in ellipse fitting to head boundary. All ellipses were properly fitted within certain margins from the head boundary without heavy outliers.
Furthermore, the proposed method effectively differentiate the head boundary patterns from the similar patterens of  the placenta boundary or uterine wall, which caused the incorrect automated caliper placement by a commercially available semiautomated program. This shows robustness of our learning-based head boundary detection.

Using the above head boundary detection, we developed an automated plane acceptance check method to determine whether the input image is acceptable for the standard plane. This uses geometric placement of three feature points (the ``box-like" CSP, the ``V-shaped" AC and Cbll) for the plane acceptance check. We conducted a feasibility test of the proposed method but with a limited training data less than 500, it was difficult to train CSP feature, especially with high variation in images. However, the test can potentially produce a high success rate of the plane acceptance check, provided sufficient amount of training data is available.

It is strongly expected that deep learning methodologies will improve their performance as training data and experience accumulate over time. The proposed method of the plane acceptance check has a room for further improvement. In order for our method to guarantee a balanced performance for the true and false cases,  a sufficient amount of training data is necessary.
The performance of acceptability scoring for CSP could be improved by reducing the search space more compactly. One may adopt the region proposal network \cite{Mask R-CNN} to get an internal module of the scoring network equipped with a kind of `attention' \cite{Attention} mechanism.

\section*{Acknowledgements}
This work was supported by the National Research Foundation of Korea (NRF) grant 2015R1A5A1009350 and 2017R1A2B20005661.

%\ifCLASSOPTIONcaptionsoff
%  \newpage
%\fi


\bibliographystyle{IEEEtran}
%\bibliography{IEEEabrv}

\end{document}